# Cost Functions over Feasible Power Transfer Regions of Virtual Power Plants

Wei Lin, *Member, IEEE*, Changhong Zhao, *Senior Member, IEEE*

*Abstract*—A virtual power plant (VPP) facilitates the integration of distributed energy resources (DERs) for the transmission-level operation. A challenge in operating a VPP is to characterize the cost function over its feasible power transfer region under DERs' uncertainties. To address this challenge, a characterization method is presented in this paper for the intraday operation of a VPP based on the concepts of nonanticipativity and robustness to DERs' uncertainties. The characterization stems from designing a second-order cone programming (SOCP) problem, based on which a feasible power transfer region across all time periods is constructed by exploring boundary points at each time period and establishing time coupling constraints. Furthermore, a cost function over the feasible power transfer region is formulated as a convex piecewise surface whose breakpoints are obtained by solving SOCP problems, together with a constant compensation cost from a linear programming problem. Finally, to alleviate the heavy computational burden brought by numerous DERs, an approximation method is presented by identifying the critical DERs whose uncertainties have dominant impacts. The effectiveness of the presented methods is verified by the numerical experiments in a 3-bus system and the IEEE 136-bus system.

## I. Introduction

*Distributed energy resources* (DERs), especially renewable resources (e.g., wind and solar), have kept growing in distribution networks [1]. To facilitate the utilization of DERs, the concept of a *virtual power plant* (VPP) has drawn much attention, leading to several practical projects (e.g., FENNIX [2], and EDISON [3]). By aggregating an entire distribution network with DERs, the VPP can participate in the transmission-level operation by adjusting its power exchanges at the *point of common coupling* (PCC) [4]. However, the operators of VPPs and transmission networks are usually different. This makes their coordination difficult in a centralized manner. As an alternative, the coordination based on the feasible power transfer region of a VPP was shown to be a promising direction [5]. The feasible power transfer region of a VPP is defined as the feasible region in the domain of PCC power exchanges (i.e., $\mathbf{P}_0^{\text{PCC}}$ and $\mathbf{Q}_0^{\text{PCC}}$ in Fig. 1). Given any such feasible power exchanges, the VPP operational constraints (e.g., power balance and voltage limits) are not violated. The feasible power transfer region can then be used as constraints in the dispatch of transmission networks [6]. Also, a VPP can bid its PCC power exchanges in a transmission electricity market [9], which motivates the formulation of a cost function over its feasible power transfer region.

Existing methods to calculate the feasible power transfer region and the corresponding cost function are mostly based on the DC power flow model. Representative methods include multi-parametric programming [7]-[8], Fourier-Motzkin elimination [9], and vertex search [10]-[11]. Specifically:

(1) Multi-parametric programming was introduced in [7] to calculate the feasible power transfer region and its cost function in a regional transmission network. The combination of active and inactive constraints in an economic dispatch problem varies with the border power flows that serve as programming parameters. For each such parameter, the Karush–Kuhn–Tucker conditions can be divided into linear equalities and inequalities. A facet of the cost function over the feasible power transfer region can be solved from the set of linear equalities, while a subset of the feasible power transfer region is specified by the inequalities. A modified method in [8] only enumerates the boundary points of the feasible power transfer region to reduce computational burden compared to [7].

(2) Fourier-Motzkin elimination in [9] projects the power flow and operational constraints onto the space of $(\mathbf{P}_0^{\text{PCC}}, \mathbf{Q}_0^{\text{PCC}})$. In that process, the variables except $(\mathbf{P}_0^{\text{PCC}}, \mathbf{Q}_0^{\text{PCC}})$ are iteratively bounded linearly in terms of other variables and thus eliminated. This method only characterizes the feasible power transfer region without providing a corresponding cost function.

(3) Vertex search was performed in [10] by solving *linear programming* (LP) problems to find boundary hyperplanes of the feasible power transfer region. Uncertainties of DERs were further incorporated in vertex search [11]. Like Fourier-Motzkin elimination, vertex search does not formulate a cost function over the feasible power transfer region.

Note that the DC power flow model is adopted in the methods mentioned above. However, the DC power flow would not be a good choice for distribution networks with significant power loss and high R/X ratios [12]. As a supplement, convex relaxations of AC power flow have drawn much attention, especially for radial distribution networks [13]-[14]. However, the nonlinearity of a relaxed AC model remains to be a major challenge to the existing methods whose foundations are the linearity in their models. Also, most existing methods neglect uncertainties of renewables, except [11] that converted uncertainties into a deterministic problem

W. Lin and C. Zhao are with the Department of Information Engineering, The Chinese University of Hong Kong, Hong Kong SAR, China. Emails: {wlin, chzhao}@ie.cuhk.edu.hk. (*Corresponding author: C. Zhao*)

This work is supported by Research Grants Council of Hong Kong through ECS Award No. 24210220 and the CUHK faculty startup grant.

by robust optimization. However, the method in [11] presets the power exchanges of a VPP in a day-ahead manner, whereas the VPP operational decisions such as generation levels can only be determined intraday after observing the realizations of uncertain DER outputs at different time periods successively. In other words, the method in [11] ignores the critical concept of nonanticipativity for stochastic cases [15]. Such neglection of nonanticipativity may jeopardize the feasibility of current decisions for future time periods [16]-[17], i.e., the robustness may not be guaranteed.

To overcome the limits mentioned above, this paper studies the cost function over the feasible power transfer region of a VPP with nonanticipativity and robustness considerations, as sketched in Fig. 1 and compared in Table 1 with previous methods. Given any pair ($\mathbf{P}_0^{PCC}$, $\mathbf{Q}_0^{PCC}$) within the feasible power transfer region in the day-ahead horizon, the intraday operation of a VPP always turns out to be feasible as the realizations of DERs' uncertainties are successively observed.

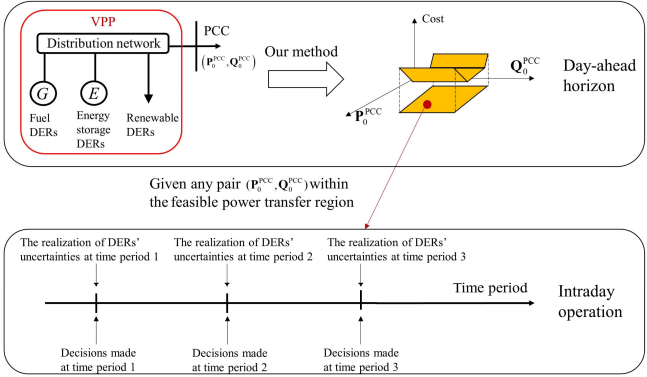

Fig. 1 A conceptual illustration of the goal in this paper. The active and reactive power exchanges at PCC are $\mathbf{P}_0^{PCC}$ and $\mathbf{Q}_0^{PCC}$, respectively.

Our major efforts to achieve such improvement are summarized as follows.

(1) A *second-order cone programming* (SOCP) problem is designed for the intraday operation of a VPP to fulfill nonanticipativity and robustness (Sec. II). The SOCP problem is designed to meet the following requirements of the intraday operation given the pre-set power exchanges in the day-ahead horizon: 1) The intraday decisions can be successively made as the realizations of DERs' uncertainties are observed, and 2) the current decisions are always feasible at this moment and at future time periods under all possible DERs' realizations.

(2) A characterization method is presented to formulate a cost function over a feasible power transfer region (Sec. III). Based on the designed SOCP problem, the feasible power transfer region across all the time periods is specified with a convex polytope, whose boundary points represent limit scenarios satisfying constraints in each time period and time-coupling constraints. Accordingly, the cost function over the feasible power transfer region is formulated as a convex piecewise surface superposed on a constant compensation cost to guarantee an upper estimation of the true cost. All the breakpoints of the convex piecewise surface are obtained by solving SOCP problems, and the constant compensation cost is obtained by solving an LP problem. To alleviate the heavy computational burden caused by numerous DERs, an approximation method is presented by identifying critical DERs whose uncertainties have dominant impacts on the VPP.

In Sec. IV, the presented methods are numerically validated in a 3-bus system and the IEEE 136-bus system. Conclusions of this paper are summarized in Sec. V.

## II. Convex Relaxation for Intraday Operation of VPP under Nonanticipativity and Robustness

Major assumptions on the VPP's operation are listed in Sec. II-A. The intraday operational requirements of a VPP are illustrated in Sec. II-B. Finally, a new SOCP problem is developed in Sec. II-C to guarantee the VPP's intraday operation to fulfill nonanticipativity and robustness.

### A. Major assumptions on the VPP's operation

Major assumptions used in this paper are listed below.

(1) Power exchanges between the VPP and the transmission network are pre-set day-ahead by minimizing the operation cost under the expected scenario (i.e., with renewable forecasts). For this, a feasible power transfer region and a cost function of the VPP are required day-ahead. During the intraday operation, the power fluctuations at PCC from the day-ahead values are limited by given thresholds [18].

(2) For illustration purposes, only fuel DERs, energy storage DERs, and renewable DERs are incorporated in the VPP. The operation cost covers fuels, charging and discharging, although other costs (e.g., for price-elastic demand) can be readily considered in our method. Particularly, fuel DERs refer to conventional units. For each unit, its cost is a convex piecewise linear function in the domain of its active generation levels [19]. For each energy storage DER, its cost to charge and discharge is a linear function [20]. Each renewable DER is represented by net active and reactive demand. In addition, the power factor at each node is assumed to be constant for convenience of discussions.

Table 1 Summary of features of different methods

| Method | Model | Feasible power transfer region | Cost function over feasible power transfer region | Nonanticipativity | Robustness |
|---|---|---|---|---|---|
| Type 1: Multi-parametric programming | DC power flow | ✓ | ✓ | ✗ | ✗ |
| Type 2: Fourier-Motzkin elimination | DC power flow | ✓ | ✗ | ✗ | ✗ |
| Type 3: Vertex search | DC power flow | ✓ | ✗ | ✗ | ✗ |
| This paper | Convex relaxation of AC power flow | ✓ | ✓ | ✓ | ✓ |

(3) DERs' uncertainties are represented by a box uncertainty set, although a budget constraint can be readily introduced to our method to filter out scenarios with low probability [21].

(4) The discussed VPP encompasses a radial distribution network where the SOCP relaxation of AC optimal power flow is exact [12]-[13].

*B. Intraday operational requirements of a VPP*

The Distflow equations [22]-[23] and SOCP relaxation [13]-[14] are employed to describe the intraday operational requirements of a VPP. The requirements presented based on whether they are coupled over time.

(1) Requirements without time coupling at time period $t$

a. Power balance at each node, i.e.,

$$P_{ij,t} - R_{ij}i_{ij,t} + \sum_{g \in G} e_{jg}P_{g,t}^G + \sum_{n \in N} e_{jn}\left(P_{n,t}^{N,d} - P_{n,t}^{N,c}\right) + e_j P_t^{PCC} \quad (1)$$
$$= \sum_{k \in L_j} P_{jk,t} + \sum_{r \in R} e_{jr}P_{r,t}^R, \forall i, \forall j, \forall t,$$

$$Q_{ij,t} - X_{ij}i_{ij,t} + \sum_{g \in G} e_{jg}Q_{g,t}^G + e_j Q_t^{PCC} \quad (2)$$
$$= \sum_{k \in L_j} Q_{jk,t} + \sum_{r \in R} e_{jr}\tan\beta_r P_{r,t}^R, \forall i, \forall j, \forall t,$$

where $P_{ij,t}$ and $Q_{ij,t}$ are active and reactive branch power flows from node $i$ to node $j$ at time period $t$, respectively; $i_{ij,t}$ is the square of current magnitude through the branch $i$-$j$; $P_{g,t}^G$ and $Q_{g,t}^G$ are active and reactive generation levels of unit $g$ at time period $t$, respectively; $P_t^{PCC}$ and $Q_t^{PCC}$ are active and reactive power exchanges at PCC at time period $t$, respectively; $P_{r,t}^R$ is net active demand at DER node $r$ at time period $t$; $P_{n,t}^{N,d}$ and $P_{n,t}^{N,c}$ are discharging and charging power of energy storage $n$ at time period $t$, respectively; $G$ is the set of generating units; $N$ is the set of energy storage nodes; $R$ is the set of renewable nodes; $L_j$ is the set of nodes that are connected to node $j$; $R_{ij}$ and $X_{ij}$ are resistance and reactance of the branch $i$-$j$; $\beta_r$ is the constant power factor angle at node $r$; $e_{jg}$, $e_{jn}$, $e_j$, and $e_{jr}$ are incident indicators.

b. The change in voltage along the branch $i$-$j$ is described below.

$$v_{j,t} = v_{i,t} - 2\left(R_{ij}P_{ij,t} + X_{ij}Q_{ij,t}\right) + \left(\left(R_{ij}\right)^2 + \left(X_{ij}\right)^2\right)i_{ij,t}, \forall i, \forall j, \forall t, \quad (3)$$

where $v_{i,t}$ is the square of voltage magnitude at node $i$ at time period $t$.

c. Flow limits. The active branch flow should not exceed its upper bound $\overline{P}_{ij}$ and lower bound $\underline{P}_{ij}$, i.e.,

$$\underline{P}_{ij} \leq P_{ij,t} \leq \overline{P}_{ij}, \forall i, \forall j, \forall t. \quad (4)$$

d. Relationship among branch flow, voltage and current, after exact SOCP relaxation [12]-[13]:

$$v_{i,t}i_{ij,t} \geq \left(P_{ij,t}\right)^2 + \left(Q_{ij,t}\right)^2, \forall i, \forall j, \forall t. \quad (5)$$

e. Generator capacities. Active generation levels should not exceed the upper bound $P_{g,t}^{Gmax}$ and lower bound $P_{g,t}^{Gmin}$, i.e.,

$$P_{g,t}^{Gmin} \leq P_{g,t}^G \leq P_{g,t}^{Gmax}, \forall g, \forall t, \quad (6)$$

where $P_{g,t}^{Gmin}$ and $P_{g,t}^{Gmax}$ are parameters selected in Appendix.

Similarly, the reactive generation levels are limited by its upper bound $\overline{Q}_g^G$ and lower bound $\underline{Q}_g^G$, i.e.,

$$\underline{Q}_g^G \leq Q_{g,t}^G \leq \overline{Q}_g^G, \forall g, \forall t, \quad (7)$$

f. Bounds of voltage magnitudes. The squared voltage magnitude at every node $i$ and time $t$ should not exceed its upper bound $\overline{v}_i$ and lower bound $\underline{v}_i$, i.e.,

$$\overline{v}_i \leq v_{i,t} \leq \underline{v}_i, \forall i, \forall t, \quad (8)$$

g. Bounds to power fluctuations at PPC. The difference between the intraday power exchanges and the day-ahead pre-set values should not exceed given thresholds, i.e.,

$$-\Delta P^{PCC} \leq P_t^{PCC} - P_{t,0}^{PCC} \leq \Delta P^{PCC}, \forall t, \quad (9)$$

$$-\Delta Q^{PCC} \leq Q_t^{PCC} - Q_{t,0}^{PCC} \leq \Delta Q^{PCC}, \forall t, \quad (10)$$

where $P_{t,0}^{PCC}$ and $Q_{t,0}^{PCC}$ are the power exchanges pre-set in the day-head horizon; $\Delta P^{PCC}$ and $\Delta Q^{PCC}$ are given thresholds.

h. Bounds of the *state of charge* (SOC). The SOC should not exceed its upper bound $S_{g,t}^{Nmax}$ and lower bound $S_{g,t}^{Nmin}$, i.e.,

$$S_{n,t}^{Nmin} \leq S_{n,t} \leq S_{n,t}^{Nmax}, \forall n, \forall t, \quad (11)$$

where $S_{n,t}$ is the SOC of energy storage $n$ at time period $t$; $S_{g,t}^{Nmin}$ and $S_{g,t}^{Nmax}$ are parameters selected in Appendix.

i. Limits of charging and discharging. Charge and discharge power $P_{n,t}^{N,c}$ and $P_{n,t}^{N,d}$ of energy storage $n$ should not exceed their upper bounds $\overline{P}_n^{N,c}$ and $\overline{P}_n^{N,d}$, respectively, i.e.,

$$0 \leq P_{n,t}^{N,c} \leq \overline{P}_n^{N,c}, \forall n, \forall t, \quad (12)$$

$$0 \leq P_{n,t}^{N,d} \leq \overline{P}_n^{N,d}, \forall n, \forall t. \quad (13)$$

j. Cyclic operation of energy storage. The SOC at the last time period $S_{n,|T|}$ should equal to the initial SOC $S_{n,\text{initial}}$, i.e.,

$$S_{n,\text{initial}} = S_{n,|T|}, \forall n. \quad (14)$$

k. Complementarity of charging and discharging. The charge and discharge of energy storage $n$ cannot simultaneously occur at time period $t$, i.e.,

$$P_{n,t}^{N,c} \times P_{n,t}^{N,d} = 0, \forall n, \forall t. \quad (15)$$

(2) Requirements with time coupling between time period $t$ and time period $t$-1

a. Ramp rates. Changes of active generation levels between consecutive periods should not exceed ramp rate limits $R_g^{down}<0$ and $R_g^{up}>0$, i.e.,

$$R_g^{down} \leq P_{g,t}^G - P_{g,t-1}^G \leq R_g^{up}, \forall g, \forall t. \quad (16)$$

b. Impacts of charging and discharging on the SOC. The SOC of energy storage $n$ is affected by its charge and discharge behaviors, as described below.

$$S_{n,t} = S_{n,t-1} + \tau_{n,c}P_{n,t}^{N,c} - P_{n,t}^{N,d}/\tau_{n,d}, \forall n, \forall t, \quad (17)$$

where $\tau_{n,c}$ and $\tau_{n,d}$ are charge and discharge efficiencies of energy storage $n$, respectively.

The intraday decisions at time period $t$ are stacked as $\mathbf{x}_t$ =$[P_t^{PCC} \ Q_t^{PCC} \ P_{g,t}^G \ Q_{g,t}^G \ P_{ij,t} \ Q_{ij,t} \ v_{i,t} \ i_{ij,t} \ S_{n,t} \ P_{n,t}^{N,c} \ P_{n,t}^{N,d}]^T \in$ $\mathbf{X}_t^{(nc)} \times \mathbf{X}_{t,t-1}^{(c)}$ where $\mathbf{X}_t^{(nc)}$ is described by (1)-(15) without time coupling, and $\mathbf{X}_{t,t-1}^{(c)}$ is described by (16)-(17) with time coupling. The uncertain quantities $P_{r,t}^R$ are confined in a box-shaped set in the following.

$$\mathbf{P}^R = \left\{ P_{r,t}^R \,\middle|\, \underline{P}_{r,t}^R \leq P_{r,t}^R \leq \overline{P}_{r,t}^R, \forall r, \forall t \right\}. \quad (18)$$

In addition, define $\mathbf{w}_t=[P_{t,0}^{\text{PCC}}\ Q_{t,0}^{\text{PCC}}]^T$ and $\mathbf{w}=[\mathbf{w}_1\ \mathbf{w}_2\ \ldots\ \mathbf{w}_{|T|}]^T$. Given a $\mathbf{w}$ in the day-ahead horizon, the decision vector $\mathbf{x}$ should be successively determined when the realizations of uncertainties in (18) are observed during the intraday horizon. This is referred to as the nonanticipativity requirement. In addition, the decision vector $\mathbf{x}_t$ at time period $t$ should be feasible at time period $t$ and at future time periods under all possible DERs' realizations. This is referred to as the robustness requirement. In the next subsection, a new SOCP problem will be presented to show existence of a nonanticipative and robust $\mathbf{w}$.

### C. SOCP problem

Our designed SOCP problem contains the constraints under the expected scenario and the constraints under the extreme scenario $s$. Particularly, $P_{r,t}^R$ equals its predicted value under the expected scenario, and $P_{r,t}^R$ under the extreme scenario $s$ is located at the $s^{th}$ vertex of the convex polytope (18). The objective of the designed problem is to minimize the operation cost under the expected scenario. The details are described below.

(1) The objective function is to minimize the total cost under the expected scenario, i.e.,

$$\min_{\substack{\{P_{g,t,0}^G,Q_{g,t,0}^G,P_{ij,t,0},Q_{ij,t,0},v_{i,t,0},S_{n,t,0},\\P_{n,t,0}^{Nc},P_{n,t,0}^{Nd},P_{t,s}^{PCC},Q_{t,s}^{PCC},P_{g,t,s}^G,Q_{g,t,s}^G,P_{ij,t,s},\\Q_{ij,t,s},v_{i,t,s},S_{n,t,s},P_{n,t,s}^{Nc},P_{n,t,s}^{Nd}\},\forall t,\forall i,\forall j,\forall s,\forall n}} \sum_{t\in T}\left(\sum_{g\in G} f_g^G\left(P_{g,t,0}^G\right)+\sum_{n\in N} c_n^{Nc} P_{n,t,0}^{Nc}+\sum_{n\in N} c_n^{Nd} P_{n,t,0}^{Nd}\right), \quad(19)$$

where $f_g^G$ is the convex piecewise linear cost function of unit $g$; $c_n^{N,c}$ and $c_n^{N,d}$ are the non-negative cost parameters of energy storage $n$ to charge and discharge, respectively; the subscript "0" indicates the variables under the expected scenario; the subscript "$s$" indicates the variables under the extreme scenario $s$.

(2) The operating region under the expected scenario at time period $t$ is the Cartesian product of the feasible set $\widetilde{\mathbf{X}}_{t,0}^{(\text{nc})}$ without time coupling and the feasible set $\widetilde{\mathbf{X}}_{t,t-1,0}^{(\text{c})}$ with time coupling. $\widetilde{\mathbf{X}}_{t,0}^{(\text{nc})}$ is defined by the constraints (1)-(14) while 1) replacing $P_{r,t}^R$ with its forecast $P_{r,t,0}^R$, and 2) replacing variables $\mathbf{x}_t$ with $\mathbf{w}_t \times \widetilde{\mathbf{x}}_{t,0}$, where $\widetilde{\mathbf{x}}_{t,0}=[P_{g,t,0}^G\ Q_{g,t,0}^G\ P_{ij,t,0}\ Q_{ij,t,0}\ v_{i,t,0}\ i_{ij,t,0}\ S_{n,t,0}\ P_{n,t,0}^{N,c}\ P_{n,t,0}^{N,d}]^T$. In addition, the following constraints are incorporated in $\widetilde{\mathbf{X}}_{t,0}^{(\text{nc})}$.

$$P_{n,t,0}^{N,c} = 0, \forall n \in \{N_{t,d} \cup N_{t,z}\}, \quad (20)$$

$$P_{n,t,0}^{N,d} = 0, \forall n \in \{N_{t,c} \cup N_{t,z}\}, \quad (21)$$

where $N_{t,c}$ is the set of energy storages that are assigned to charge at time period $t$; $N_{t,d}$ is the set of energy storages that are assigned to discharge at time period $t$; $N_{t,z}$ is the set of energy storages that are assigned not to charge or discharge at time period $t$. The sets $N_{t,c}$, $N_{t,d}$ and $N_{t,z}$ are determined in Appendix. In particular, they are mutually exclusive and satisfy:

$$N_{t,d} \cup N_{t,c} \cup N_{t,z} = N. \quad (22)$$

$\widetilde{\mathbf{X}}_{t,t-1,0}^{(\text{c})}$ is defined by the following constraints:

$$S_{n,t,0} = S_{n,t-1,0} + \tau_{n,c} P_{n,t,0}^{N,c}, \forall n \in N_{t,c}, \quad (23)$$

$$S_{n,t,0} = S_{n,t-1,0} - P_{n,t,0}^{N,d}/\tau_{n,d}, \forall n \in N_{t,d}, \quad (24)$$

$$S_{n,t,0} = S_{n,t-1,0}, \forall n \in N_{t,z}. \quad (25)$$

Note that $\widetilde{\mathbf{X}}_{t,0}^{(\text{nc})}$ and $\widetilde{\mathbf{X}}_{t,t-1,0}^{(\text{c})}$ are in the domain of $\widetilde{\mathbf{x}}_{t,0}$ when $\mathbf{w}_t$ is given as parameters.

(3) The operation region under the extreme scenario $s$ at time period $t$ is the Cartesian product of $\mathbf{X}_{t,s}^{(\text{nc})}$ and $\mathbf{X}_{t,t-1,s}^{(\text{c})}$. $\mathbf{X}_{t,t-1,s}^{(\text{c})}$ is defined by the constraints (23)-(25) while replacing $\widetilde{\mathbf{x}}_{t,0}$ with $\mathbf{x}_{t,s}=[P_{t,s}^{\text{PCC}}\ Q_{t,s}^{\text{PCC}}\ P_{g,t,s}^G\ Q_{g,t,s}^G\ P_{ij,t,s}\ Q_{ij,t,s}\ v_{i,t,s}\ i_{ij,t,s}\ S_{n,t,s}\ P_{n,t,s}^{N,c}\ P_{n,t,s}^{N,d}]^T$. $\mathbf{X}_{t,s}^{(\text{nc})}$ is defined by the constraints (1)-(14) and (20)-(21) while 1) replacing $\mathbf{x}_t$ with $\mathbf{x}_{t,s}$, and 2) replacing $P_{r,t}^R$ with $P_{r,t,s}^R$, i.e., the realization of $P_{r,t}^R$ at the $s^{th}$ vertex of (18). In addition, the following two addition constraints are involved in $\mathbf{X}_{t,s}^{(\text{nc})}$ for every pair of scenarios $n,s \in S$.

$$v_{i,t,s} i_{ij,t,n} \geq (P_{ij,t,s})^2 + (P_{ij,t,n})^2, \forall i, \forall j, \forall t, \forall n, \forall s, \quad (26)$$

$$v_{i,t,n} i_{ij,t,s} \geq (Q_{ij,t,n})^2 + (Q_{ij,t,s})^2, \forall i, \forall j, \forall t, \forall n, \forall s. \quad (27)$$

Note that the parameters $P_{g,t}^{G\min}$, $P_{g,t}^{G\max}$, $S_{n,t}^{N\min}$, and $S_{n,t}^{N\max}$ selected in Appendix guarantee existence of the rules described below.

$$\underline{P}_g^G \leq P_{g,t}^{G\min} \leq P_{g,t}^{G\max} \leq \overline{P}_g^G, \forall t, \forall g, \quad (28)$$

$$R_g^{\text{down}} \leq P_{g,t}^{G\min} - P_{g,t-1}^{G\max}, \forall t, \forall g, \quad (29)$$

$$P_{g,t}^{G\max} - P_{g,t-1}^{G\min} \leq R_g^{\text{up}}, \forall t, \forall g, \quad (30)$$

$$\underline{S}_n^N \leq S_{n,t}^{N\min} \leq S_{n,t}^{N\max} \leq \overline{S}_n^N, \forall t, \forall n, \quad (31)$$

$$\underline{P}_{n,t}^{N,c} \leq \left(S_{n,t}^{N\min} - S_{n,t-1}^{N\max}\right)/\tau_{n,c}, \forall t, \forall n \in N_{t,c}, \quad (32)$$

$$\left(S_{n,t}^{N\max} - S_{n,t-1}^{N\min}\right)/\tau_{n,c} \leq \overline{P}_{n,t}^{N,c}, \forall t, \forall n \in N_{t,c}, \quad (33)$$

$$\underline{P}_{n,t}^{N,d} \leq \tau_{n,d}\left(S_{n,t-1}^{N\min} - S_{n,t}^{N\max}\right), \forall t, \forall n \in N_{t,d}, \quad (34)$$

$$\tau_{n,d}\left(S_{n,t-1}^{N\max} - S_{n,t}^{N\min}\right) \leq \overline{P}_{n,t}^{N,d}, \forall t, \forall n \in N_{t,d}, \quad (35)$$

where $\overline{P}_g^G$ and $\underline{P}_g^G$ are maximum and minimum physical capacities of unit $g$, respectively; $\overline{S}_n^N$ and $\underline{S}_n^N$ are maximum and minimum physical capacities of energy storage $n$, respectively.

For convenience of discussions, a separable version of the SOCP problem designed above is given in a compact form as follows:

$$z = \min_{\{\widetilde{\mathbf{x}}_{t,0},\mathbf{x}_{t,s}\},\forall t,\forall s} \sum_t f_t(\widetilde{\mathbf{x}}_{t,0}), \quad (36)$$

s.t. $\mathbf{w}_t \times \widetilde{\mathbf{x}}_{t,0} \in \widetilde{\mathbf{X}}_{t,0}^{(\text{nc})} \times \widetilde{\mathbf{X}}_{t,t-1,0}^{(\text{c})}, \mathbf{x}_{t,s} \in \mathbf{X}_{t,s}^{(\text{nc})} \times \mathbf{X}_{t,t-1,s}^{(\text{c})}, \forall t, \forall s, (37)$

where $\sum_t f_t(\widetilde{\mathbf{x}}_{t,0})$ is the cost in (19).

For our SOCP problem designed in (36)-(37), a proposition and its proof are provided below.

**Proposition 1**. Given $\mathbf{w}=[\mathbf{w}_1\ \mathbf{w}_2\ \ldots\ \mathbf{w}_{|T|}]^T$ with which the SOCP problem (36)-(37) is feasible, there is a nonanticipative and robust intraday decision vector $\mathbf{x}_t=[P_t^{\text{PCC}}\ Q_t^{\text{PCC}}\ P_{g,t}^G\ Q_{g,t}^G\ P_{ij,t}$ $Q_{ij,t}\ v_{i,t}\ i_{ij,t}\ S_{n,t}\ P_{n,t}^{N,c}\ P_{n,t}^{N,d}]^T \in \mathbf{X}_t^{(\text{nc})} \times \mathbf{X}_{t,t-1}^{(\text{c})}$. ∎

**Proof**. The vector $\mathbf{x}_t$ can be determined by the following rules:

$$\mathbf{x}_t = \sum_{s \in S} \lambda_{t,s} \mathbf{x}_{t,s}, \tag{38}$$

where $\lambda_{t,s}$ is calculated as follows from the realization $P_{r,t}^{R,o}$ of $P_{r,t}^{R}$ observed at time period $t$:

$$P_{r,t}^{R,o} = \sum_{s \in S} \lambda_{t,s} P_{r,t,s}^{R}, \forall r, \forall t, \tag{39}$$

$$\sum_{s \in S} \lambda_{t,s} = 1, \forall t, \tag{40}$$

$$\sum_{s \in S} \lambda_{t,s} = 1, \lambda_{t,s} \geq 0, \forall s, \forall t, \tag{41}$$

Next, it is proven that that a nonanticipative and robust decision $\mathbf{x}_t \in \mathbf{X}_t^{(nc)} \times \mathbf{X}_{t,t-1}^{(c)}$ can be made.

(1) Nonanticipativity: The decision $\mathbf{x}_t$ can be made at each time $t$ as (38), which only relies on the realization of $P_{r,t}^{R}$ at time period $t$.

(2) The proof of robustness lies in proving that $\mathbf{x}_t$ from (38) satisfies the constraints (1)-(17).

a. Fulfillment of (1)-(4) and (6)-(14). The nonnegative parameters $\lambda_{t,s}$ in (39)-(41) are multiplied on both sides of the constraint that corresponds to the constraint (1) in $\mathbf{X}_{t,s}^{(nc)}$. Adding them up yields

$$\sum_{s \in S}(\lambda_{t,s} P_{ij,t,s}) - \sum_{s \in S}(\lambda_{t,s} R_{ij} i_{ij,t,s}) + \sum_{s \in S}\left(\lambda_{t,s} \sum_{g \in G} e_{jg} P_{g,t,s}^{G}\right) + \sum_{s \in S}\left(\lambda_{t,s} \sum_{n \in N} e_{jn}\left(P_{n,t,s}^{N,d} - P_{n,t,s}^{N,c}\right)\right)$$
$$+ \sum_{s \in S}\left(\lambda_{t,s} e_j P_{t,s}^{PCC}\right) = \sum_{s \in S}\left(\lambda_{t,s} \sum_{k \in L_j} P_{jk,t,s}\right) + \sum_{s \in S}\left(\lambda_{t,s} \sum_{r \in R} e_{jr} P_{r,t,s}^{R,o}\right), \forall i, \forall j. \tag{42}$$

which by (38)-(41) becomes

$$P_{ij,t} - R_{ij} i_{ij,t} + \sum_{g \in G} e_{jg} P_{g,t}^{G} + \sum_{n \in N} e_{jn}\left(P_{n,t}^{N,d} - P_{n,t}^{N,c}\right) + e_j P_t^{PCC}$$
$$= \sum_{k \in L_j} P_{jk,t} + \sum_{r \in R} e_{jr} P_{r,t}^{R,o}, \forall i, \forall j. \tag{43}$$

By (43), the constraint (1) is satisfied given the realization $P_{r,t}^{R,o}$ of DER output. The satisfaction of (2)-(4) and (6)-(14) can be proven in a similar way.

b. Fulfillment of (5). Factors $(\lambda_{t,s})^2$ are multiplied on both sides of the constraint that corresponds to (5) in $\mathbf{X}_{t,s}^{(nc)}$. Adding them up yields

$$\sum_{s \in S}\left((\lambda_{t,s})^2 v_{i,t,s} i_{ij,t,s}\right) \geq \sum_{s \in S}\left((\lambda_{t,s})^2 (P_{ij,t,s})^2\right) + \sum_{s \in S}\left((\lambda_{t,s})^2 (Q_{ij,t,s})^2\right), \forall i, \forall j. \tag{44}$$

Factors $\lambda_{t,s}\lambda_{t,n}$ for all pairs $(s,n) \in S^2, s \neq n$ are multiplied on both sides of (26). Adding them up yields

$$\sum_{(s,n) \in S^2, s \neq n}(\lambda_{t,s}\lambda_{t,n} v_{i,t,s} i_{ij,t,n}) \geq \sum_{(s,n) \in S^2, s \neq n}\left(\lambda_{t,s}\lambda_{t,n}(P_{ij,t,s})^2\right) + \sum_{(s,n) \in S^2, s \neq n}\left(\lambda_{t,s}\lambda_{t,n}(P_{ij,t,n})^2\right), \forall i, \forall j.$$
$$\Rightarrow \sum_{(s,n) \in S^2, s \neq n}(\lambda_{t,s}\lambda_{t,n} v_{i,t,s} i_{ij,t,n}) \geq 2 \sum_{(s,n) \in S^2, s \neq n}(\lambda_{t,s}\lambda_{t,n} P_{ij,t,s} P_{ij,t,n}), \forall i, \forall j. \tag{45}$$

Similar to the derivation of (45), the following inequality can be derived based on (27):

$$\sum_{(s,n) \in S^2, s \neq n}(\lambda_{t,s}\lambda_{t,n} v_{i,t,s} i_{ij,t,n}) \geq 2 \sum_{(s,n) \in S^2, s \neq n}(\lambda_{t,s}\lambda_{t,n} Q_{ij,t,s} Q_{ij,t,n}), \forall i, \forall j. \tag{46}$$

Adding (44)-(46) yields

$$\sum_{s \in S}\left((\lambda_{t,s})^2 v_{i,t,s} i_{ij,t,s}\right) + \sum_{(s,n) \in S^2, s \neq n}(\lambda_{t,s}\lambda_{t,n} v_{i,t,s} i_{ij,t,n}) + \sum_{(s,n) \in S^2, s \neq n}(\lambda_{t,s}\lambda_{t,n} v_{i,t,s} i_{ij,t,n})$$
$$\geq \sum_{s \in S}\left((\lambda_{t,s})^2 (P_{ij,t,s})^2\right) + \sum_{s \in S}\left((\lambda_{t,s})^2 (Q_{ij,t,s})^2\right) \tag{47}$$
$$+ 2\sum_{(s,n) \in S^2, s \neq n}(\lambda_{t,s}\lambda_{t,n} P_{ij,t,s} P_{ij,t,n}) + 2\sum_{(s,n) \in S^2, s \neq n}(\lambda_{t,s}\lambda_{t,n} Q_{ij,t,s} Q_{ij,t,n}), \forall i, \forall j.$$

The inequality (47) can be reorganized as follows:

$$\left(\sum_{s \in S} \lambda_{t,s} v_{i,t,s}\right) \times \left(\sum_{s \in S} \lambda_{t,s} i_{ij,t,s}\right) \geq \left(\sum_{s \in S} \lambda_{t,s} P_{ij,t,s}\right)^2 + \left(\sum_{s \in S} \lambda_{t,s} Q_{ij,t,s}\right)^2, \forall i, \forall j. \tag{48}$$

Combining (38) and (48) leads to satisfaction of (5).

c. Fulfillment of (15). The complementarity can be satisfied from the separation of energy storage based on the three sets $N_{t,c}$, $N_{t,d}$, $N_{t,z}$ that are mutually exclusive.

d. Fulfillment of (16). The nonnegative parameters $\lambda_{t,s}$ in (39)-(41) are multiplied on both sides of the constraint that corresponds to (6) in $\mathbf{X}_{t,s}^{(nc)}$. This leads to

$$\lambda_{t,s} P_{g,t}^{Gmin} \leq \lambda_{t,s} P_{g,t,s}^{G} \leq \lambda_{t,s} P_{g,t}^{Gmax}, \forall t, \forall g. \tag{49}$$

which implies:

$$\lambda_{t,s}\left(P_{g,t}^{Gmin} - P_{g,t-1}^{Gmax}\right) \leq \lambda_{t,s}\left(P_{g,t,s}^{G} - P_{g,t-1,s}^{G}\right) \leq \lambda_{t,s}\left(P_{g,t}^{Gmax} - P_{g,t-1}^{Gmin}\right), \forall t, \forall g, \forall s. \tag{50}$$

Adding (50) across all extreme scenarios up yields

$$\left(P_{g,t-1}^{Gmin} - P_{g,t}^{Gmax}\right) \leq \left(P_{g,t}^{G} - P_{g,t-1}^{G}\right) \leq \left(P_{g,t-1}^{Gmax} - P_{g,t}^{Gmin}\right), \forall t, \forall g. \tag{51}$$

Note that the selection of $P_{g,t}^{Gmin}$, $P_{g,t-1}^{Gmax}$, $P_{g,t}^{Gmax}$ and $P_{g,t-1}^{Gmin}$ satisfies (29)-(30). This leads to

$$R_g^{down} \leq \left(P_{g,t}^{G} - P_{g,t-1}^{G}\right) \leq R_g^{up}, \forall t, \forall g, \tag{52}$$

i.e., the constraint (16) holds.

e. Fulfillment of (17) can be proven based on the constraints that correspond to (23)-(25) in $\mathbf{X}_{t,t-1,s}^{(c)}$, in a similar way to the fulfillment of (1)-(4) and (6)-(14). Note that the $(S_{n,t}, P_{n,t}^{N,c}, P_{n,t}^{N,d})$ from (17) also satisfies under the parameter selection conditions (32)-(35). ∎

**Remark 1**. Proposition 1 and its proof can be readily extended to the case with budget constraints, because the intersection of the uncertainty set (18) and the budget-constrained set is still a convex set, such that (39)-(41) can be still applied to the proof of constraint fulfillment. ∎

The SOCP problem in (36)-(37) is formulated for each given $\mathbf{w} = [P_{t,0}^{PCC} \ Q_{t,0}^{PCC}]^T \forall t$. Any such $\mathbf{w}$ that makes (1)-(17) feasible under nonanticipativity and robustness is called a feasible power transfer of the VPP, and the minimum objective value (19) can be treated as a cost function of $\mathbf{w}$. It is critical for a VPP to characterize its feasible power transfer region and cost function in the domain of $\mathbf{w}$, from which the transmission operator can solve for its day-ahead dispatch [11]. This characterization will be elaborated in next section.

## III. CALCULATION OF COST FUNCTION OVER THE FEASIBLE POWER TRANSFER REGION OF VPP

The exploration of the feasible power transfer region will be explained in Sec. IV-A, followed by the modeling of the cost function in Sec. IV-B. To alleviate the computational burden, an approximation method is introduced in Sec. IV-C.

## A. Exploration of the feasible power transfer region

The formal definition of a feasible power transfer region is:

$$\Omega \triangleq \left\{ \mathbf{w} \left| \begin{array}{l} \exists \left(\tilde{\mathbf{x}}_{t,0}, \mathbf{x}_{t,s}\right) \text{ such that } \mathbf{w}_t \times \tilde{\mathbf{x}}_{t,0} \in \tilde{\mathbf{X}}_{t,0}^{(\text{nc})} \times \tilde{\mathbf{X}}_{t,t-1,0}^{(\text{c})} \\ \text{and } \mathbf{x}_{t,s} \in \mathbf{X}_{t,s}^{(\text{nc})} \times \mathbf{X}_{t,t-1,s}^{(\text{c})}, \forall t, \forall s \end{array} \right. \right\}. \quad (53)$$

The feasible power transfer region $\Omega$ defined in (53) is in the domain of $\mathbf{w}$ whose dimension is rather high when multiple time periods are involved. The calculation of a high-dimension polytope usually faces a considerable computational burden [25]-[26]. Consequently, a successive determination strategy is presented in this subsection for a fast calculation. The key idea of the presented strategy lies in 1) calculating the feasible power transfer region $\Omega_{t-1}$ at time period $t$-1 by finding the vertices of the constraints at time period $t$-1, 2) incorporating the time coupling constraints given $\Omega_{t-1}$ to connect the constraints at time period $t$. Finally, the intersection of feasible power transfer regions at all the time periods is employed as an estimation of $\Omega$.

The produce above is elaborated as follows.

(1) Calculation of $\Omega_{t-1}$

The feasible power transfer region $\Omega_{t-1}$ at time period $t$-1 is mathematically defined below.

$$\Omega_{t-1} \triangleq \left\{ \mathbf{w}_{t-1} \left| \begin{array}{l} \exists \mathbf{w}_{t-1} \times \left(\tilde{\mathbf{x}}_{t-1,0}, \boldsymbol{\mu}_{t-2}\right) \in \tilde{\mathbf{X}}_{t-1,0}^{(\text{nc})} \times \tilde{\mathbf{X}}_{t-1,t-2,0}^{(\text{cnew})}, \\ \exists \left(\mathbf{x}_{t-1,s}, \boldsymbol{\mu}_{t-2}\right) \in \mathbf{X}_{t-1,s}^{(\text{nc})} \times \mathbf{X}_{t-1,t-2,s}^{(\text{cnew})}, \forall s \end{array} \right. \right\}, \quad (54)$$

where $\boldsymbol{\mu}_{t-2}$ is the vector of ancillary variables in real space; $\tilde{\mathbf{X}}_{t-1,t-2,0}^{(\text{cnew})}$ and $\mathbf{X}_{t-1,t-2,s}^{(\text{cnew})}$ are convex feasible sets that reflect the time coupling of $\mathbf{X}_{t-1,t-2,0}^{(\text{c})}$ and $\mathbf{X}_{t-1,t-2,s}^{(\text{c})}$, respectively. The explicit construction of $\tilde{\mathbf{X}}_{t-1,t-2,0}^{(\text{cnew})}$ and $\mathbf{X}_{t-1,t-2,s}^{(\text{cnew})}$ will be found in (59).

Considering the convexity of $\tilde{\mathbf{X}}_{t-1,0}^{(\text{nc})} \times \tilde{\mathbf{X}}_{t-1,t-2,0}^{(\text{cnew})}$ and $\mathbf{X}_{t-1,s}^{(\text{nc})} \times \mathbf{X}_{t-1,t-2,s}^{(\text{cnew})}$, the set $\Omega_{t-1}$ can be computed using the following vertex search method:

Step 1: Initialization. SOCP problems are solved subject to $\tilde{\mathbf{X}}_{t-1,0}^{(\text{nc})} \times \tilde{\mathbf{X}}_{t-1,t-2,0}^{(\text{cnew})}$ and $\mathbf{X}_{t-1,s}^{(\text{nc})} \times \mathbf{X}_{t-1,t-2,s}^{(\text{cnew})}$ for all scenarios $s$ to obtain the minimum and maximum of each element in $\mathbf{w}_{t-1}$. Collect those minimum and maximum points in a set $V_0$.

Step 2: Polytope construction. A polytope $\mathbf{R}$ is constructed with vertices $V_0$, using tools such as MPT3 [24]. The half-space representation of the polytope $\mathbf{R}$ is $\mathbf{A}_{t-1}\mathbf{w}_{t-1} \leq \mathbf{B}_{t-1}$.

Step 3: Exploration. For each facet of the polytope $\mathbf{R}$, the following SOCP problem is solved:

$$\max_{\{\mathbf{w}_{t-1}, \tilde{\mathbf{x}}_{t-1,0}, \mathbf{x}_{t-1,s}, \boldsymbol{\mu}_{t-2}\}, \forall s} \mathbf{A}_{t-1,j}\mathbf{w}_{t-1}, \quad (55)$$

$$\text{s.t.} \quad \mathbf{w}_{t-1} \times \left(\tilde{\mathbf{x}}_{t-1,0}, \boldsymbol{\mu}_{t-2}\right) \in \tilde{\mathbf{X}}_{t-1,0}^{(\text{nc})} \times \tilde{\mathbf{X}}_{t-1,t-2,0}^{(\text{cnew})}, \quad (56)$$

$$\left(\mathbf{x}_{t-1,s}, \boldsymbol{\mu}_{t-2}\right) \in \mathbf{X}_{t-1,s}^{(\text{nc})} \times \mathbf{X}_{t-1,t-2,s}^{(\text{cnew})}, \forall s, \quad (57)$$

where $\mathbf{A}_{t-1,j}$ is the $j^{th}$-row of $\mathbf{A}_{t-1}$.

Intuitively, (55)-(57) move the facet $\mathbf{A}_{t-1,j}\mathbf{w}_{t-1}=\mathbf{B}_{t-1,j}$ as far as possible away from the center of the current feasible region $\Omega_{t-1}$. After solving (55)-(57) for all rows $j$, collect all the optimal $\mathbf{w}_{t-1}$ in a set $V_{\text{new}}$, such that $\{V_{\text{new}} \cup V_0\}$ replaces $V_0$ as the vertices of a new polytope $\mathbf{R}_{\text{new}}$.

Step 4: Termination check. Check the difference between the volumes of $\mathbf{R}$ and $\mathbf{R}_{\text{new}}$ [25]. Once the difference is smaller than a given threshold, terminate the algorithm and return $\mathbf{R}_{\text{new}}$ as $\Omega_{t-1}$; otherwise, let $\mathbf{R}_{\text{new}} \rightarrow \mathbf{R}$ and go back to Step 3.

(2) Explicit construction of $\tilde{\mathbf{X}}_{t,t-1,0}^{(\text{cnew})}$ and $\mathbf{X}_{t,t-1,s}^{(\text{cnew})}$

Observed from (53), $\Omega$ is tightly coupled across all the time periods by $\mathbf{X}_{t,t-1,0}^{(\text{c})}$ and $\mathbf{X}_{t,t-1,s}^{(\text{c})}$ for all scenarios $s$. In $\mathbf{X}_{t,t-1,0}^{(\text{c})}$ and $\mathbf{X}_{t,t-1,s}^{(\text{c})}$, time coupling arises from the state of charge changes. The state of charge changes between two consecutive time periods can be presented in the following compact form:

$$\mathbf{U}_t^{\text{SOC}}\mathbf{y}_t = \mathbf{U}_{t-1}^{\text{SOC}}\mathbf{y}_{t-1} + \mathbf{U}_t^{\text{P}}\mathbf{y}_t, \quad (58)$$

where $\mathbf{y}_t = \begin{bmatrix} \tilde{\mathbf{x}}_{t,0}^T & \mathbf{x}_{t,1}^T & \dots & \mathbf{x}_{t,|S|}^T \end{bmatrix}^T$; $\mathbf{U}_t^{\text{SOC}}$ are constant matrices to extract variables of the state of charge at time period $t$ from $\mathbf{y}_t$; $\mathbf{U}_t^{\text{P}}$ are constant matrices to extract charge and discharge variables at time period $t$ from $\mathbf{y}_t$. Note that the constraint (58) represent the changes of state of charge in (17).

Motivated by the convexity of $\tilde{\mathbf{X}}_{t-1,0}^{(\text{nc})} \times \tilde{\mathbf{X}}_{t,t-1,0}^{(\text{c})}$ and $\tilde{\mathbf{X}}_{t,s}^{(\text{nc})} \times \tilde{\mathbf{X}}_{t,t-1,s}^{(\text{c})}$ for all scenarios $s$, $\tilde{\mathbf{X}}_{t,t-1,0}^{(\text{cnew})}$ and $\mathbf{X}_{t,t-1,s}^{(\text{cnew})}$ are explicitly constructed as follows:

$$\mathbf{U}_t^{\text{SOC}}\mathbf{y}_t = \mathbf{U}_{t-1}^{\text{SOC}} \sum_{j \in J_{t-1}} \mu_{t-1}^{(j)}\mathbf{y}_{t-1}^{(j)} + \mathbf{U}_t^{\text{P}}\mathbf{y}_t, \sum_{j \in J_{t-1}} \mu_{t-1}^{(j)} = 1, \mu_{t-1}^{(j)} \geq 0, \quad (59)$$

where $\mathbf{y}_{t-1}^{(j)}$ is the optimal solution of $\mathbf{y}_{t-1}$ that corresponds to the $j^{th}$ vertex of $\Omega_{t-1}$ obtained in Steps 1-4.

Once the strategy above is implemented, $\Omega$ defined in (53) can be estimated by $\tilde{\Omega}$ in the following.

$$\tilde{\Omega} \triangleq \left\{ (\mathbf{w}, \boldsymbol{\mu}) \left| \begin{array}{l} \mathbf{w}_t = \sum_{j \in J_t} \mu_t^{(j)} \mathbf{w}_t^{(j)}, \forall t, \\ \mathbf{U}_{t-1}^{\text{SOC}} \sum_{j \in J_{t-1}} \mu_{t-1}^{(j)}\mathbf{y}_{t-1}^{(j)} = \mathbf{U}_t^{\text{SOC}} \sum_{j \in J_t} \mu_t^{(j)}\mathbf{y}_t^{(j)}, \forall t, \\ \sum_{j \in J_t} \mu_t^{(j)} = 1, \mu_t^{(j)} \geq 0, \forall t \end{array} \right. \right\}, \quad (60)$$

where $\boldsymbol{\mu}$ is a vector that is composed of $\boldsymbol{\mu}_t = \begin{bmatrix} \mu_t^{(1)} & \dots & \mu_t^{(|J_t|)} \end{bmatrix}^T$ across all time periods. Particularly, a property of $\tilde{\Omega}$ to guarantee feasibility is provided below.

**Proposition 2.** $\tilde{\Omega}$ is an inner estimation of $\Omega$. ∎

**Proof.** Denote a vector as follows:

$$\mathbf{x}^{\text{whole}} = \begin{bmatrix} \mathbf{w}_1^T & \dots & \mathbf{w}_{|T|}^T & \left(\mathbf{U}_1^{\text{SOC}}\mathbf{y}_1\right)^T & \dots & \left(\mathbf{U}_{|T|}^{\text{SOC}}\mathbf{y}_{|T|}\right)^T \end{bmatrix}^T. \quad (61)$$

The constraint (37) can be projected onto the space of $\mathbf{x}^{\text{whole}}$. Due to the convexity of (37), this projection leads to a convex polytope. Some vertices of this projection are found based on Steps 1-4 and (59). These vertices can be further projected onto the space of $\mathbf{w}$ to get $\tilde{\Omega}$. As a result, $\tilde{\Omega}$ is an inner estimation. ∎

## B. Calculation of the cost function

A cost function at each time period is required for VPP bidding [27]. However, time coupling in (36)-(37) leads to difficulties for individually formulating the cost function at each time period. This issue will be resolved by calculating the cost function at each time period. Their summation superposed on a compensation cost is above the true total cost function. Our method is elaborated below.

(1) Calculation of a cost function $z_t(\mathbf{w}_t)$ at time period $t$

Considering the following optimization problem to minimize the cost at time period $t$:

$$z_t(\mathbf{w}_t) = \min_{\{g, \tilde{\mathbf{x}}_{t,0}, \mathbf{x}_{t,s}\}, \forall t, \forall s} g, \quad (62)$$

$$\text{s.t.} \quad f_t(\tilde{\mathbf{x}}_{t,0}) \leq g \leq f_t(\tilde{\mathbf{x}}_{t,0}^{\max}) \text{ and (37),} \quad (63)$$

where $\tilde{\mathbf{x}}_{t,0}^{\max}$ is the upper bound of $\tilde{\mathbf{x}}_{t,0}$; $g$ is a cost variable in real space [28].

Note that $\mathbf{w}_t$ is involved in (37) as coverage parameters. This is why the minimum (62) can be expressed as a function of $\mathbf{w}_t$.

The convexity of $z_t(\mathbf{w}_t)$ is provided below, as a foundation of our method.

**Proposition 3**. $z_t(\mathbf{w}_t)$ is convex in the domain of $\mathbf{w}_t$. ∎

**Proof**. Due to the convexity of (63), the projection of (63) onto the space of $(g,\mathbf{w}_t)$ is a convex polytope. Hence, $(g,\mathbf{w}_t)$ can be a convex combination of vertices, i.e.,

$$\begin{bmatrix} g & \mathbf{w}_t^T \end{bmatrix}^T = \sum_{k \in K_t} \alpha_t^{(k)} \mathbf{V}^{(k)}, \alpha_t^{(k)} \geq 0, \sum_{k \in K_t} \alpha_t^{(k)} = 1, \quad (64)$$

where $\alpha_t^{(k)}$ is the coefficient associated with the $k^{th}$ vertex $\mathbf{V}^{(k)}$.

Based on (64) and Gaussian elimination, $g$ can be formulated as a linear function of $\mathbf{w}_t$ and $\boldsymbol{\alpha}_t$, where $\boldsymbol{\alpha}_t$ are coefficients in (64) that are not eliminated. The linear function is denoted as $g = G_t(\mathbf{w}_t, \boldsymbol{\alpha}_t)$. As a result, $z_t(\mathbf{w}_t)$ can be formulated in the following.

$$z_t(\mathbf{w}_t) = \sup_{\boldsymbol{\alpha}_t}(-G_t(\mathbf{w}_t, \boldsymbol{\alpha}_t)). \quad (65)$$

Eq. 3.7 in [29] tells the convexity of piecewise maximum of linear functions (65), i.e., $z_t(\mathbf{w}_t)$ is convex in the domain of $\mathbf{w}_t$. ∎

Motivated by the convexity illustrated in Proposition 3, the following auxiliary region $\boldsymbol{\Psi}_t$ is constructed to calculate the cost function $z_t(\mathbf{w}_t)$ over the feasible power transfer region $\tilde{\boldsymbol{\Omega}}$:

$$\boldsymbol{\Psi}_t \triangleq \left\{ (g, \mathbf{w}, \boldsymbol{\mu},) \cup \{\tilde{\mathbf{x}}_{t,0}, \mathbf{x}_{t,s}\}, \forall t, \forall s \, \middle| \, \begin{array}{c} \text{Constraint (62),} \\ (\mathbf{w}, \boldsymbol{\mu}) \in \tilde{\boldsymbol{\Omega}} \end{array} \right\}. \quad (66)$$

Let $\boldsymbol{\Phi}_t$ denote the projection of $\boldsymbol{\Psi}_t$ to the domain of $(g_t, \mathbf{w}_t)$. The values of $g_t$ at some boundary points of $\boldsymbol{\Phi}_t$ are $z_t(\mathbf{w}_t)$. Consequently, the remaining work is to identify the right facets from $\boldsymbol{\Phi}_t$ whose intersections are those boundary points.

For the $j^{th}$ facet $g_t = \mathbf{C}_{tj}\mathbf{w}_t + E_{tj}$ in $\boldsymbol{\Phi}_t$, a point $\mathbf{w}_{tj}$ located on this facet can be easily found. Given $\mathbf{w}_{tj}$, the following SOCP problem is solved:

$$\min_{\{g, \tilde{\mathbf{x}}_{t,0}, \mathbf{x}_{t,s}\}, \forall t, \forall s} g, \text{ s.t. } \mathbf{w}_t = \mathbf{w}_{tj} \text{ and (37).} \quad (67)$$

Denote the minimum objective value of (67) as $g_t^*$. If $\mathbf{C}_{tj}\mathbf{w}_{tj} + E_{tj} = g_t^*$, the $j^{th}$ facet of $\boldsymbol{\Phi}_t$ attains $z_t(\mathbf{w}_t)$, otherwise (if $\mathbf{C}_{tj}\mathbf{w}_{tj} + E_{tj} > g_t^*$) the infimum is not attained. Fig. 2 illustrates these two cases.

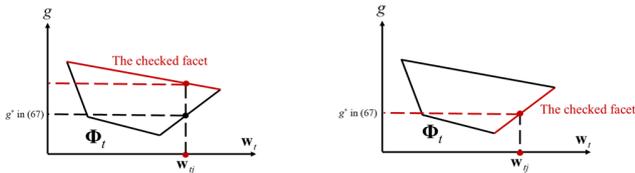

(a) The infimum is not attained     (b) The infimum is attained

Fig. 2 Two cases regarding a facet of $\boldsymbol{\Phi}_t$ checked through (67)

(2) Calculation of a compensation cost $\varepsilon$

If the same $\mathbf{w}$ is given in (36)-(37) and (62)-(63), the summation of $z_t(\mathbf{w}_t)$ in (62)-(63) across all the time periods will not exceed the true total cost $z$ in (36)-(37) across all the time periods. Therefore, a compensation cost $\varepsilon$ is necessary to guarantee the coverage of the true total cost when $z_t(\mathbf{w}_t)$ is employed for bidding. In other words, such a desired $\varepsilon$ should guarantee

$$\sum_{t \in T} z_t(\mathbf{w}_t) + \varepsilon \geq z, \forall (\mathbf{w}, \boldsymbol{\mu}) \in \tilde{\boldsymbol{\Omega}}. \quad (68)$$

To achieve this goal, an LP problem is designed as follows:

$$\max_{\{\mathbf{w}, \boldsymbol{\mu}, \tilde{\mathbf{x}}_{t,0}, \mathbf{x}_{t,s}\} \forall t, \forall s} \left( \sum_{t \in T} f_t^{\text{new}}(\tilde{\mathbf{x}}_{t,0}) - \sum_{t \in T} z_t(\mathbf{w}_t) \right), \quad (69)$$

$$\text{s.t.} \quad (\mathbf{w}, \boldsymbol{\mu}) \in \tilde{\boldsymbol{\Omega}} \text{ and (37),} \quad (70)$$

where $f_t^{\text{new}}$ is the same function $f_t$ in (36) except that the cost function $f_g$ in $f_t$ is revised as a linear function by directly connecting the beginning point and the endpoint of $f_t$.

**Proposition 4**. The optimal objective value of (69)-(70) can be taken as $\varepsilon$ that satisfies (68).

**Proof**. Consider the following bi-level optimization problem:

$$\max_{(\mathbf{w}, \boldsymbol{\mu})} \left( \left( z = \min_{\{\tilde{\mathbf{x}}_{t,0}, \mathbf{x}_{t,s}\}, \forall t, \forall s} \sum_{t \in T} f_t(\tilde{\mathbf{x}}_{t,0}) \right) - \sum_{t \in T} z_t(\mathbf{w}_t) \right) \text{ s.t. } (\mathbf{w}, \boldsymbol{\mu}) \in \tilde{\boldsymbol{\Omega}}. \quad (71)$$

The optimal objective value of (71) is the maximum error between the summation of $z_t(\mathbf{w}_t)$ across time periods and the true total cost $z$. This yields the satisfaction of (68) when the optimal objective value of (71) is taken as $\varepsilon$.

Revising the inner optimization problem of (71) as a maximum problem leads to

$$\max_{(\mathbf{w}, \boldsymbol{\mu})} \left( \left( z = \max_{\{\tilde{\mathbf{x}}_{t,0}, \mathbf{x}_{t,s}\}, \forall t, \forall s} \sum_{t \in T} f_t(\tilde{\mathbf{x}}_{t,0}) \right) - \sum_{t \in T} z_t(\mathbf{w}_t) \right) \text{ s.t. } (\mathbf{w}, \boldsymbol{\mu}) \in \tilde{\boldsymbol{\Omega}}. \quad (72)$$

Based on [30], the bi-level problem in (72) is equivalent to the following single-level problem:

$$\max_{\{\mathbf{w}, \boldsymbol{\mu}, \tilde{\mathbf{x}}_{t,0}, \mathbf{x}_{t,s}\}, \forall t, \forall s} \left( \sum_{t \in T} f_t(\tilde{\mathbf{x}}_{t,0}) - \sum_{t \in T} z_t(\mathbf{w}_t) \right) \text{ s.t. } (\mathbf{w}, \boldsymbol{\mu}) \in \tilde{\boldsymbol{\Omega}} \text{ and (37).} \quad (73)$$

The optimal objective value of (73) is larger than that of (71); thus, (68) holds if the optimal objective value of (73) is taken as $\varepsilon$. If $f_t$ in (73) is replaced with the function $f_t^{\text{new}}$ in (69), (73) becomes the LP problem in (69)-(70). Note that the function $f_t^{\text{new}}$ in (69) is always above the function $f_t$. Consequently, the optimal objective function of (69)-(70) is larger than that of (73). Consequently, Proposition 4 holds. ∎

**Remark 2**. The method above also works for other convex cost functions $f_t$, such as the costs for elastic demand response. ∎

*C. Fast approximation method by identifying critical DERs*

The calculation of the cost function over the feasible power transfer region may suffer a heavy computational burden brought by numerous DERs in a VPP. This issue arises from repeatedly solving SOCP problems for all the $2^M$ extreme scenarios of the box uncertainty set across M DERs.

To alleviate computational burden, we propose a heuristic approximation method, where the impacts of DERs'

uncertainties are ranked and only the DERs with the most significant impacts are considered in the extreme scenarios.

In particular, DERs' uncertainties just appear in power balance constraints (1)-(2). The impact of a DER lies in its fluctuation interval [$\underline{P}_{r,t}^{R}$, $\overline{P}_{r,t}^{R}$] and its coefficients in (1) and (2), i.e., (1+tan $\beta_r$). Therefore, we rank (1+tan $\beta_r$)×($\overline{P}_{r,t}^{R}$ - $\underline{P}_{r,t}^{R}$) over DERs $r$ to assess their impact at time $t$.

IV. CASE STUDIES

The presented methods are verified in a 3-bus system and the IEEE 136-bus test system [31]. In the 3-bus system, DERs' uncertainties are introduced at node 2 and node 3. In the IEEE 136-bus test system, uncertainties of DERs occur at node 8, node 41, node 44, node 45, and node 106.

Firstly, the feasible power transfer region obtained by our method is verified to guarantee nonanticipativity and robustness in the 3-bus system over two time periods, as will be shown in Sec. V-A. Secondly, the cost function obtained by our method is verified to guarantee the coverage of the true costs in the 3-bus system, as will be shown in Sec V-B. Finally, the effectiveness of our approximation method is verified in the IEEE 136-bus system, as will be shown in Sec. V-C. All numerical results are calculated with MATLAB R2012a and performed on a laptop equipped with Intel (R) Core (TM) i7-8565U CPU @ 1.80GHz 8.00G RAM. All SOCP problems and LP problems are solved via YALMIP and CPLEX.

*A. Validation of nonanticipativity and robustness in the feasible power transfer region*

In this subsection, it will be shown that any pair ($\mathbf{P}_0^{PCC}$, $\mathbf{Q}_0^{PCC}$) selected from our feasible power transfer region can guarantee the nonanticipative and robust intraday operation of a VPP. The 3-bus system over two time periods [31] are considered. Also, a conventional unit is installed at node 1 and an energy storage is installed at node 2.

To verify nonanticipativity, the following procedure is implemented: 1) 10000 points in our feasible power transfer region are randomly sampled; 2) for each sample, a realization of DERs' uncertainties at time period 1 is randomly selected from the uncertainty set (18); 3) Given the uncertainty realization at time period 1, a feasible decision vector $\mathbf{x}_1$ is obtained by solving the optimization problem whose objective function is set as zero and constraints are (1)-(17); 4) given $\mathbf{x}_1$ and the uncertainty realization at time period 2, a feasible decision vector $\mathbf{x}_2$ is obtained by solving the optimization problem whose objective function is set as zero and constraints are (1)-(17).

Changes of active generation between time periods 1 and 2 are calculated and shown in Fig. 3. It shows that the ramp rate constraints are satisfied, which validates nonanticipativity of the intraday decisions, as claimed in Proposition 1.

In addition, the energy storage charging power at time period 1 and discharging power at time period 2 are shown in Fig. 4. The limits of charge and discharge rates are respected even if decisions of energy storage are successively made as time goes. This follows Proposition 1 proven in Sec. III-C.

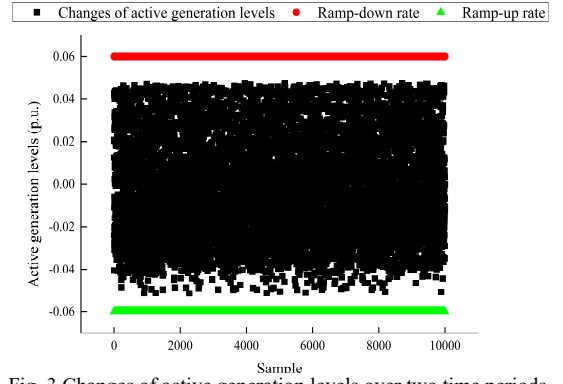
Fig. 3 Changes of active generation levels over two time periods.

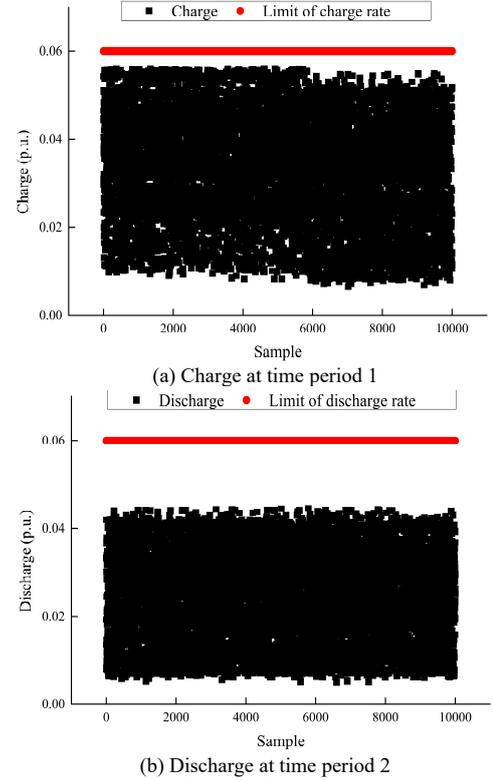
(a) Charge at time period 1

(b) Discharge at time period 2
Fig. 4 Behaviors of energy storage

Furthermore, for the previous 10000 samples selected from the feasible power transfer region, ten realizations of DERs' uncertainties are randomly selected from the uncertainty set (18). For each sample and each realization, a decision vector $\mathbf{x}$ is solved by the optimization problem whose objective function is set as zero and constraints are (1)-(17). Our result is that each sample in the feasible power transfer region with each realization of DER uncertainties leads to a feasible decision, thus validating robustness.

*B. Validation of the cost function*

Our cost function over the feasible power transfer region is examined in the previous 3-bus system. As illustrated in Sec. III-B, our cost function is composed of two parts. One is functions $z_1(P_{1,0}^{PCC}, Q_{1,0}^{PCC})$ and $z_2(P_{2,0}^{PCC}, Q_{2,0}^{PCC})$ at time periods 1 and 2, and the other one is the compensation cost $\varepsilon$. The compensation cost $\varepsilon$ is obtained as \$ 300.3 by solving (69)-(70), and the functions $z_1$ and $z_2$ are plotted in Fig. 5.

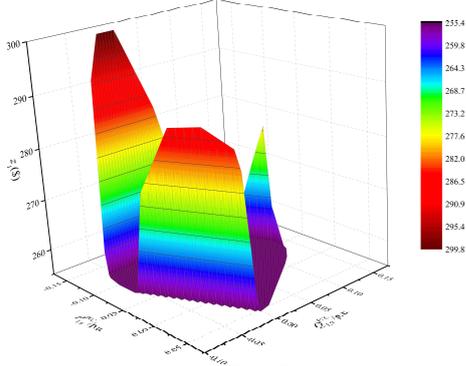

(a) The function $z_1(P_{1,0}^{PCC}, Q_{1,0}^{PCC})$ at time period 1

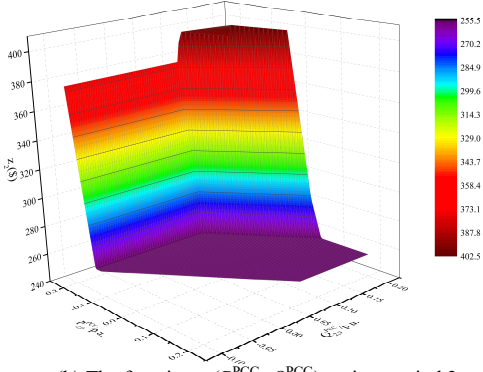

(b) The function $z_2(P_{2,0}^{PCC}, Q_{2,0}^{PCC})$ at time period 2

Fig. 5 Cost functions at two time periods

Furthermore, the bid of the VPP is set as our cost function over the feasible power transfer region, i.e., the bid is $z_1(P_{1,0}^{PCC}, Q_{1,0}^{PCC}) + z_2(P_{2,0}^{PCC}, Q_{2,0}^{PCC}) + \varepsilon$. For each of the 10000 samples, our bids are compared with the true total costs that are obtained by solving the SOCP problem (36)-(37). The numerical results are shown in Fig. 6. Our bids are always above the true total costs. Consequently, our method can guarantee to cover the true total cost with the cost function for the bid at each time period.

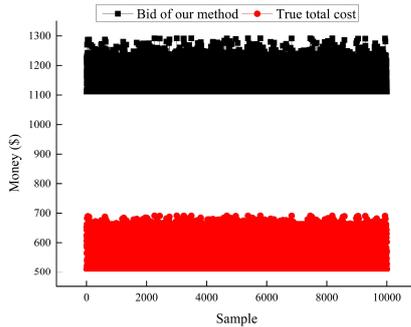

Fig. 6 Analysis of absolution errors of costs

### C. Validation of the approximation method

Consider the IEEE 136-bus system with five DERs over one time period, in which there are $2^5=32$ extreme scenarios that require heavy computation. We evaluate the following cases as computationally efficient approximations:

*Case 1*: Consider expected values of DER outputs at all the nodes.

*Case 2*: Consider only the uncertainties at node 106 and expected values at all the other nodes.

*Case 3*: Consider only the uncertainties at node 44 and expected values at all the other nodes.

*Case 4*: Consider only the uncertainties at node 8 and expected values at all the other nodes.

Using the criterion in Sec. III-C, the uncertainties at node 106 make the most significant impact and those at node 8 are most insignificant. Computation time and volumes of the feasible power transfer regions are compared over four cases in Table 2.

Table 2 Comparison of the four cases with different uncertainties considered.

| Case | Computational Time (Second) | Volume (MW×MVar) |
| --- | --- | --- |
| 1 | 8.669 | 38.649 |
| 2 | 20.095 | 36.999 |
| 3 | 15.594 | 37.976 |
| 4 | 17.262 | 38.648 |

Case 1 (all expected values) needs the shortest computation time and returns the coarsest approximation of the feasible power transfer region. Case 2, which incorporates the uncertainties making the most significant impact, achieves a more stringent feasible region with moderate increase in computation time.

## V. CONCLUSIONS

To facilitate the participation of a *virtual power plant* (VPP) into the transmission-level operation, a novel SOCP problem and a characterization method are presented to determine the cost function over the feasible power transfer region of a VPP under uncertainties of DERs. Nonanticipativity and robustness in operating a VPP can be guaranteed based on the presented model and method. Also, a heuristic is presented to select the DERs with dominant impacts to alleviate the heavy computational burden associated with a large number of uncertain scenarios. The accuracy and computational efficiency of the presented methods were validated through numerical experiments in a 3-bus system and the IEEE 136-bus system.

## APPENDIX

The parameters $P_{g,t}^{Gmin}$, $P_{g,t}^{Gmax}$, $S_{n,t}^{Nmin}$, and $S_{n,t}^{Nmax}$ used in (36)-(37) can be determined by solving the following SOCP problem:

$$\max_{\{\tilde{\mathbf{x}}_{t,0}, \mathbf{w}_t, P_{g,t}^{Gmin}, P_{g,t}^{Gmax}, S_{n,t}^{Nmin}, S_{n,t}^{Nmax}\}, \forall t, \forall g} \left\{ \sum_{t \in T} \left[ \sum_{g \in G} \left( \frac{P_{g,t}^{Gmax} - P_{g,t}^{Gmin}}{\overline{P}_g^G - \underline{P}_g^G} \right) + \sum_{n \in N} \left( \frac{S_{n,t}^{Nmax} - S_{n,t}^{Nmin}}{\overline{S}_n^N - \underline{S}_n^N} \right) - \xi \left( \sum_{n \in N} P_{n,t}^{Nd} + P_{n,t}^{Nc} \right) \right] \right\}, \quad (A1)$$

$$\text{s.t.} \quad (\tilde{\mathbf{x}}_{t,0}, \mathbf{w}_t) \in \mathbf{X}_{t,0}^A, \forall t, \quad (A2)$$

$$(28)\text{-}(31), \quad (A3)$$

$$S_{n,t-1}^{max} - S_{n,t}^{min} \leq \overline{P}_{n,t}^{N,d} / \tau_{n,d}, \forall t, \forall n, \quad (A4)$$

$$S_{n,t}^{max} - S_{n,t-1}^{min} \leq \tau_{n,c} \overline{P}_{n,t}^{N,c}, \forall t, \forall n, \quad (A5)$$

where $\xi$ is a given small positive number; the feasible set $\mathbf{X}_{t,0}^A$ is established by the constraints (1)-(14) and (16)-(17) while 1) replacing $P_{r,t}^R$ with its forecast $P_{r,t,0}^R$, and 2) replacing variables $\mathbf{x}_t$ with $\mathbf{w}_t \times \tilde{\mathbf{x}}_{t,0}$.

Solving (A1)-(A5) provides an opportunity to get a solution where the complementarity constraint (15) holds. This occurs when the dual variables associated with power balance in $\mathbf{X}_{t,0}^A$ are non-negative. This proof is like the proof in Sec. III-C in [32]. If the dual variables obtained by (A1)-(A5) are negative, the complementarity constraint (15) will not hold. Under such as case, (A1)-(A5) will be revised by 1) deleting the final item in (A1) and 2) introducing binary variables to guarantee the complementarity constraint (15) [33]. A solution can be obtained by solving the revised (A1)-(A5) to determine the parameters $P_{g,t}^{\text{Gmin}}$, $P_{g,t}^{\text{Gmax}}$, $S_{n,t}^{\text{Nmin}}$, and $S_{n,t}^{\text{Nmax}}$.

Once the parameters $P_{g,t}^{\text{Gmin}}$, $P_{g,t}^{\text{Gmax}}$, $S_{n,t}^{\text{Nmin}}$, and $S_{n,t}^{\text{Nmax}}$ are determined, the sets $N_{t,c}$, $N_{t,d}$, and $N_{t,z}$ are further determined based on the solution. The set $N_{t,c}$ includes energy storages which are charging at time period $t$. The set $N_{t,d}$ includes energy storages which are discharging at time period $t$. Other energy storages that are neither $N_{t,c}$ nor $N_{t,d}$ belong to $N_{t,z}$.